\documentclass[a4paper,
               biblatex,     
               ]{jacow}
\usepackage{pdfpages,multirow,ragged2e} %
\usepackage{wasysym} 
\usepackage{xcolor}
\makeatletter%
	\ifboolexpr{bool{xetex}}
	 {\renewcommand{\Gin@extensions}{.pdf,%
	                    .png,.jpg,.bmp,.pict,.tif,.psd,.mac,.sga,.tga,.gif,%
	                    .eps,.ps,%
	                    }}{}
\makeatother

\ifboolexpr{bool{xetex} or bool{luatex}} 
 {}                                      
 {\usepackage[utf8]{inputenc}}           

\usepackage[USenglish]{babel}

\ifboolexpr{bool{jacowbiblatex}}%
 {%
  \addbibresource{WEPM120.bib}
}{}

\begin{document}

\title{Evaluation of a High-Power Target Design for Positron Production at CEBAF\thanks{This work was supported by the European Union’s Horizon 2020 Research and Innovation program under Grant Agreement No 824093 and the U.S. DOE, Office of Science, Office of Nuclear Physics, contract DE-AC05-06OR23177.}}

\author{A. Ushakov\textsuperscript{1,2}\thanks{ushakov@jlab.org}, 
S. Covrig\textsuperscript{2}, 
J. Grames\textsuperscript{2}, 
S. Habet\textsuperscript{1,2}, 
C. Le Galliard\textsuperscript{1}, 
E. Voutier\textsuperscript{1} \\
		\textsuperscript{1}Université Paris-Saclay, Orsay, France \\
		\textsuperscript{2}Thomas Jefferson National Accelerator Facility, Newport News, VA, USA}
	
\maketitle

\begin{abstract}
A source for polarized positron beams at the Continuous Electron Beam Accelerator Facility (CEBAF) at Jefferson Lab is being designed. The Polarized Electrons for Polarized Positrons (PEPPo) concept is used to produce polarized e$^+$e$^-$-pairs from the bremsstrahlung radiation of a longitudinally polarized electron beam interacting within a high-$Z$ conversion target. The scheme under consideration includes a 4~mm thick tungsten target that absorbs 17~kW deposited by a 1~mA continuous-wave electron beam with an energy of 120~MeV. The concept of a rotating tungsten rim mounted on a water-cooled copper disk was explored. The results of ANSYS thermal and mechanical analyses are discussed together with FLUKA evaluations of the radiation damages.
\end{abstract}

\section{INTRODUCTION}

Positron beams (polarized and unpolarized) are an option for potential upgrades of CEBAF~\cite{arrington2022physics}. 
Conducting measurements with a  positron beam will provide new experimental observables and will subsequently expand the physics reach of Jefferson Lab~\cite{Accardi_2021}. The PEPPo experiment \cite{PhysRevLett.116.214801} has demonstrated high efficiency of polarization transfer from electrons to the positrons through a two-step process: bremsstrahlung followed by pair production, with both reactions taking place (in series) in the same physical target. 
The polarization transfer is almost 100\% at the high end of the positron energy spectrum. The positron polarization is proportional to the energy, but the number of highly polarized positrons is inversely proportional to the energy. Therefore, the quantity of interest, which characterizes a polarized source and enters the statistical error of the measurement of experimental signals sensitive to the beam polarization, is the Figure-of-Merit (FoM) corresponding to the product of the beam current $I$ with the square of the average longitudinal polarization $\overline{P_z}$ of the beam population (FoM = $I \overline{P_z}^2$). The FoM was used to optimize the target thickness and to select the positron energies~\cite{habet:2023} caught by the capture system of the polarized positron injector~\cite{Habet:2022fch}. The essential differences between PEPPo and conventional unpolarized positron sources are using an initially polarized electron beam and selecting high-energy positron slices, an energy region featuring high polarization transfer~\cite{Olsen:1959zz,Kuraev:2009uy}.

Jefferson Lab plans to repurpose the Low Energy Recirculation Facility (LERF) for the generation, capture and acceleration of positron beams up to 123~MeV. A polarized electron source produces a continuous-wave (CW) high current (>1~mA) and high polarization (>90\%) beam, which is accelerated up to 120~MeV towards a high-power target for positron production. The optimal thickness of the tungsten target at 120~MeV is 4~mm~\cite{habet:2023}. A significant fraction of beam power is deposited in the target. 
The high non-uniform power deposition, quick temperature rise, mechanical stress and radiation damage may cause target failure. 
The present work discusses a first evaluation of the possible parameters of the high-power target like the energy deposition, the level of radiation damage, and the expected operational temperature and mechanical stress. 

\section{Energy Deposition and Target Design}

\begin{figure}[!b]
	\centering
	\includegraphics*[width=82mm]{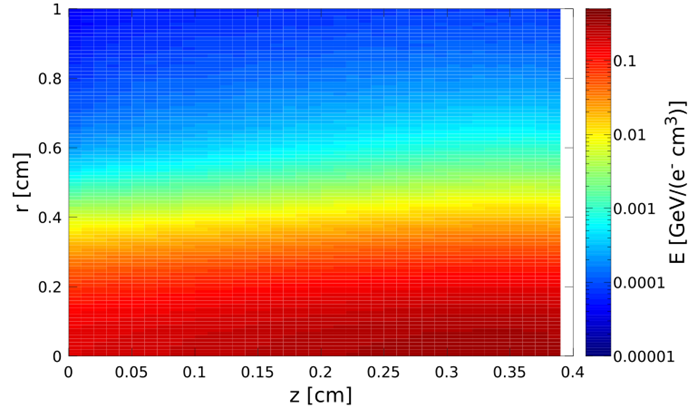}
	\caption{Energy deposition profile of a 120~MeV electron beam with a 1.5~mm RMS spot size in a 4~mm thick tungsten target.}
	\label{fig:Edep-2D-FLUKA}
\end{figure}

The energy deposition of the electron beam in the target was determined with FLUKA~\cite{FLUKA1:ref,FLUKA2:ref}. The distribution of the energy deposited by a 120~MeV electron beam with a 1.5~mm RMS spot size in a stationary tungsten target of 4~mm thickness is shown in Fig.~\ref{fig:Edep-2D-FLUKA}.
The FLUKA data on the deposited energy was converted into power and imported as a heat source into ANSYS~\cite{ANSYSCFD:ref} to determine the temperature profile. For the 1~mA CW beam, the peak power density of 324~W/mm$^3$ corresponds to 324~MeV/(e$^-\cdot$mm$^3$).

\begin{figure}[!t]
	\centering
	\includegraphics*[width=73.0mm]{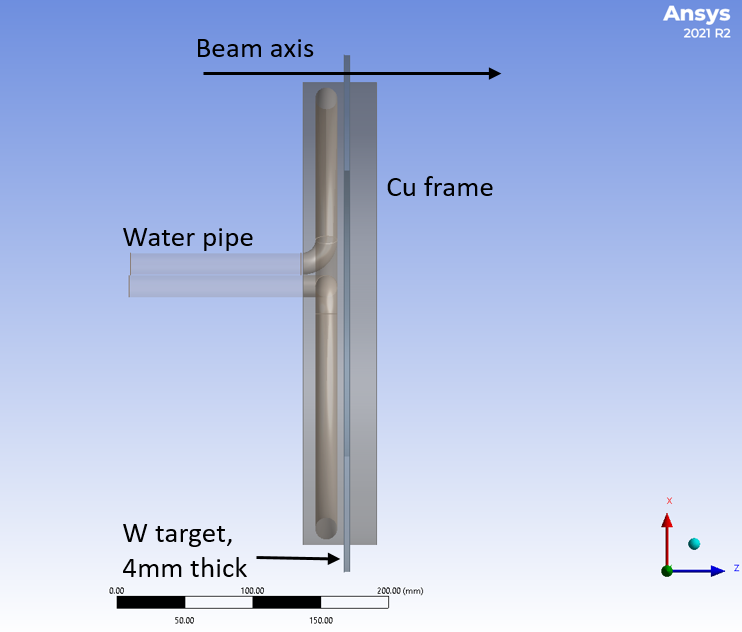}
	\caption{Conceptual design of the rotating target (side view).}
	\label{fig:Target-Concept}
\end{figure}

To keep the temperature of the tungsten target at an acceptable level, the heat generated in the target by the beam must be distributed over a larger volume. The tungsten rim with a thickness of 4~mm is mounted on a water-cooled copper disk.
The side-view of the considered conceptual design of rotating target is shown in Fig.~\ref{fig:Target-Concept}.
The outer radius of the tungsten rim is 19~cm. The beam passes the target at a radius of 18~cm.
The target rotation frequency considered in the temperature calculations was 2 Hz and the tangential speed of the beam moving on target was 2.3~m/s.
The water channel inside the copper disk has a radius of 8 mm.

\section{Temperature and Stress Calculations}

Time-dependent CFD simulations were implemented.
ANSYS Fluent has been used to calculate the temperature profile in the target material.
The water flowing with a speed of 1.5~m/s (0.3~kg/s mass flow, 10~kPa pressure loss between the inlet and outlet) was cooling the copper disk and tungsten rim mounted on the copper disk.  
For the 17~kW deposited in target beam power, the estimated maximal temperature of the water was about 30$^{\circ}$C, and the peak temperature in the copper disk was below 100$^{\circ}$C.

\begin{figure}[tb]
	\centering
	\includegraphics*[width=78mm]{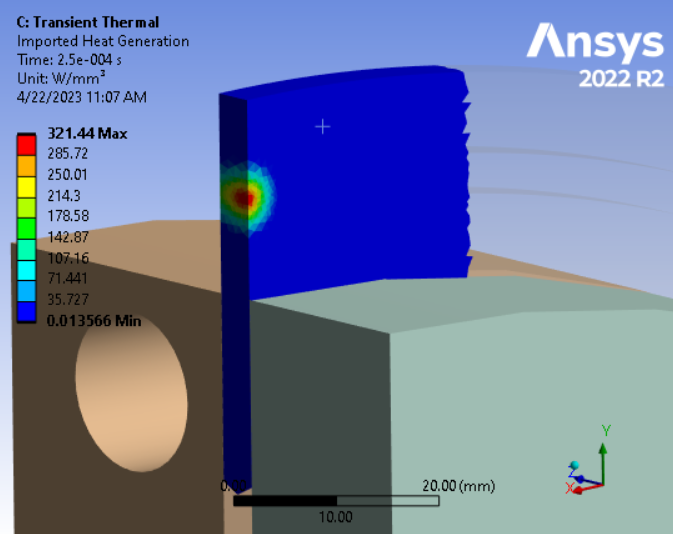}
	\caption{Distribution of heat power density of 1 mA electron beam at 120 MeV and 1.5 mm rms size in 4 mm thick tungsten.}
	\label{fig:ImportedHeat}
\end{figure}

\begin{figure}[!t]
	\centering
	\includegraphics*[width=76mm]{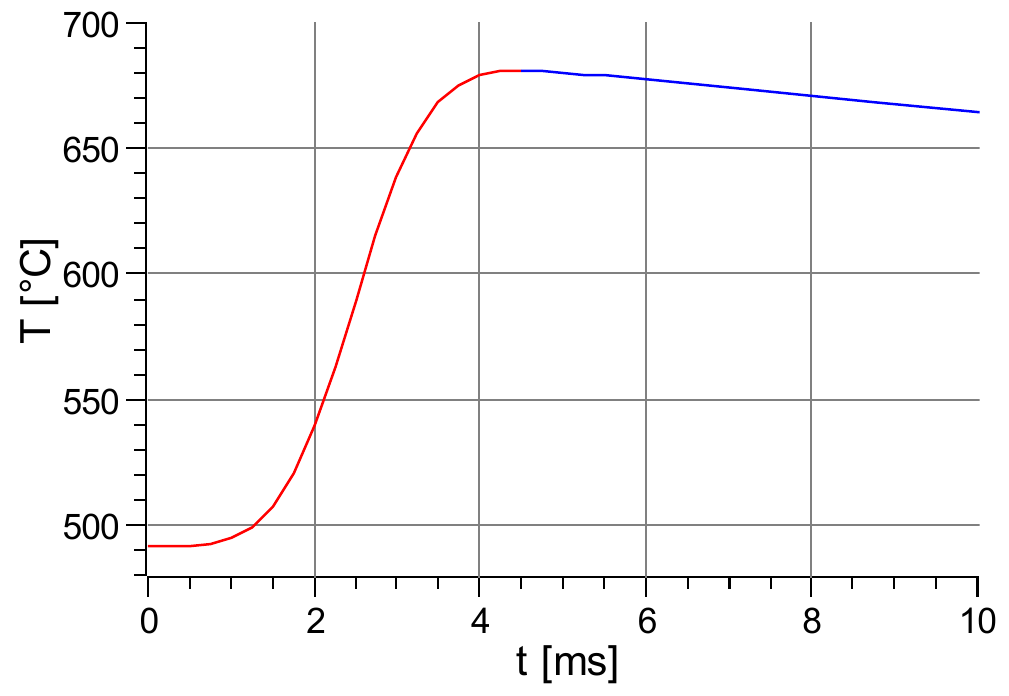}
	\caption{Cycling temperature in tungsten at radius of 18~cm rotated with the 2.3 m/s tangential velocity and 1.5 mm RMS beam spot (heating phase is shown in red and beginning cooling phase continued up to 0.5~s in blue).}
	\label{fig:Transient-Cycle}
\end{figure}

\begin{figure}[!t]
	\centering
	\includegraphics*[width=78mm]{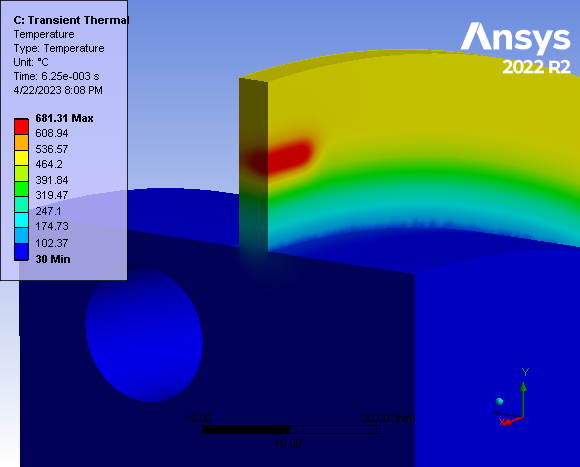}
	\caption{Temperature distribution in 17~kW target.}
	\label{fig:Tmax}
\end{figure}

To simulate the heating of tungsten by the electron beam moving on the target, the distribution of heat power density was shifted by 0.56 mm along the circular path with a radius of 18~cm in 0.25 ms time steps (2.3 m/s). The heat distribution at one time step is shown in Fig.~\ref{fig:ImportedHeat}. For the selected point on target (at $R = 18$~cm) and RMS beam size of 1.5 mm, the temperature rises during 4.5~ms and reaches the maximum of 681$^{\circ}$C. Figure~\ref{fig:Transient-Cycle} shows the time evolution of temperature during the first 10~ms of the 0.5~s cycle (one complete turn of target). The spatial  distribution of temperature is shown in Fig.~\ref{fig:Tmax}.

\begin{figure}[!t]
	\centering
	\includegraphics*[width=78mm]{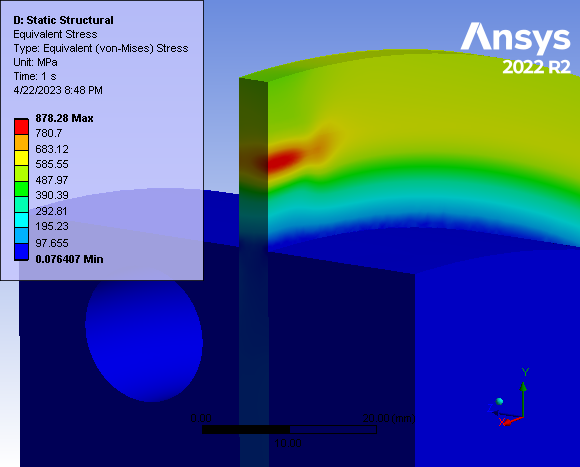}
	\caption{Mechanical stress induced by the temperature.}
	\label{fig:EqStress}
\end{figure}

The temperature distribution was imported into ANSYS static structural module to calculate the mechanical stress. Von Mises stress is a good measure of the proximity to failure of a material with values below material yield stress indicating an elastic behaviour~\cite{Stein:2001jdc}. Figure~\ref{fig:EqStress} shows the spatial distribution of equivalent von Mises stress. The maximal stress in tungsten is 878~MPa. Experimental testing of tungsten at these levels of mechanical stress and temperature is planned.   

\section{Radiation Damage}

\begin{figure}[!htb]
	\centering
	\includegraphics*[width=76mm]{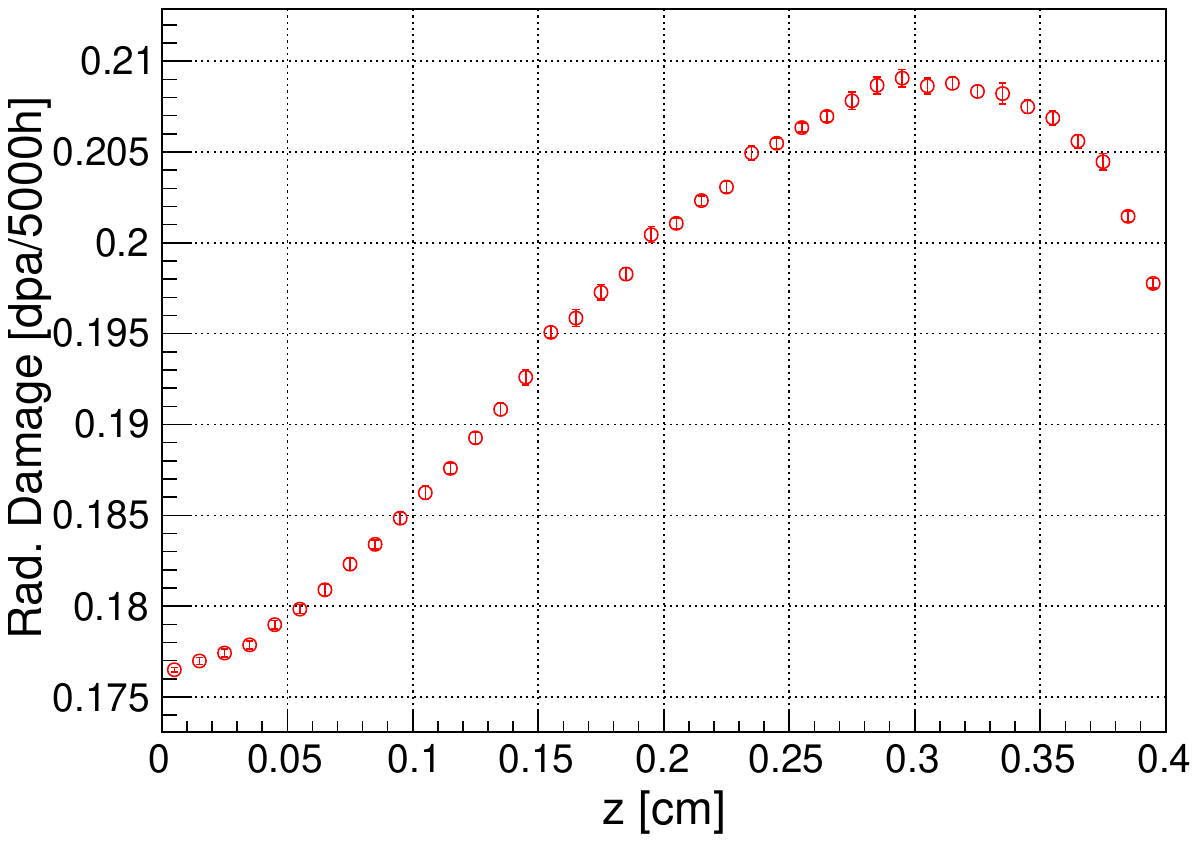}
	\caption{Radiation damage of W-target with $\diameter 36$~cm.}
	\label{fig:dpa-sum-vs-Z-vD35cm.pdf}
\end{figure}

FLUKA is used to determine the radiation damage of tungsten. Radiation damage effects are implemented in FLUKA for all particles, including recoils which have enough energy to induce damage to the materials~\cite{FLUKA1:ref}. The maximal damage is $5.7\cdot10^{-22}$ displacements per atom (dpa) per e$^-$. Figure~\ref{fig:dpa-sum-vs-Z-vD35cm.pdf} shows the radiation damage at different depths of the spinning target with a diameter of 36~cm after 5000~h of irradiation. The calculated peak damage is 0.21~dpa. The effect of such radiation damage on material properties should be experimentally verified.

\section{OUTLOOK}

A high-power target for positron production at the future Ce$^+$BAF positron injector~\cite{ipac23:grames} was evaluated. 
Time-dependent CFD simulations (ANSYS Fluent) were implemented.
The temperature, mechanical stress and radiation damage were calculated for the tungsten target with a thickness of 4 mm and 17~kW power deposited by 1~mA CW electron beam with an energy of 120~MeV. 
The peak temperature of the target rotated with a velocity of 2.3~m/s is 680~°C and the maximal equivalent von Mises stress is 880~MPa. The estimated annual radiation damage is 0.21~dpa. 
To check if the target can be used safely over an extended period under such conditions and find experimentally the endurance stress limits and the impact of radiation damage on the material properties, tests of target materials (tungsten and tantalum) using \SI{50}{\micro A} at 3.5~MeV electron beam at Mainz Microtron (MAMI) have been started. Material fatigue tests using a high power laser are also planned, which will be similar to performed tungsten foil tests for the APEX target at Jefferson Lab \cite{APEXtarget}.

%
%
\ifboolexpr{bool{jacowbiblatex}}%
	{\printbibliography}%
	{%
	
 } 
			
\end{document}